\documentclass[prb,aps,superscriptaddress,floatfix,tightenlines,showpacs,twocolumn]{revtex4-2}
\usepackage{graphicx}
\usepackage{dcolumn}   % needed for some tables
\usepackage{bm}        % for math
\usepackage{amssymb}   % for math
\usepackage{amsmath}
\usepackage{url}
\usepackage{layout}
\usepackage{color}
\usepackage{epsfig}
\usepackage{graphicx}
\usepackage{upgreek}
\usepackage{booktabs}
\usepackage{float}
\usepackage[hyperindex,breaklinks]{hyperref}
\def\be{\begin{equation}}       \def\ee{\end{equation}}
\def\bea{\begin{eqnarray}}      \def\eea{\end{eqnarray}}
\def\bp{\begin{pmatrix}} \def\ep{\end{pmatrix}}
\def\beaa{\begin{equation}\begin{aligned}}
\def\eeaa{\end{aligned}\end{equation}}

\begin{document}

\title{Frustrated superconductivity and sextetting order}

\author{Zhiming Pan}
\thanks{These two authors contributed equally to this work.}
\affiliation{Institute for Theoretical Sciences, Westlake University, Hangzhou 310024, Zhejiang, China}
\affiliation{New Cornerstone Science Laboratory, Department of Physics, School of Science, Westlake University, Hangzhou 310024, Zhejiang, China}
\author{Chen Lu}
\thanks{These two authors contributed equally to this work.}
\affiliation{New Cornerstone Science Laboratory, Department of Physics, School of Science, Westlake University, Hangzhou 310024, Zhejiang, China}
\author{Fan Yang}
\email{yangfan\_blg@bit.edu.cn}
\affiliation{School of Physics, Beijing Institute of Technology, Beijing 100081, China}
\author{Congjun Wu}
\email{wucongjun@westlake.edu.cn}
\affiliation{New Cornerstone Science Laboratory, Department of Physics, School of Science, Westlake University, Hangzhou 310024, Zhejiang, China}
\affiliation{Institute for Theoretical Sciences, Westlake University, Hangzhou 310024, Zhejiang, China}
\affiliation{Key Laboratory for Quantum Materials of Zhejiang Province, School of Science, Westlake University, Hangzhou 310024, Zhejiang, China}
\affiliation{Institute of Natural Sciences, Westlake Institute for Advanced Study, Hangzhou 310024, Zhejiang, China}

%\date{\today}

\begin{abstract}
The superconducting state typically favors a uniform spatial distribution akin to ferromagnetism. 
Nevertheless, the pair-density-wave state exhibits sign changes
in the pairing order, leading to potential frustrations in phase coherence.
We propose a mechanism to the sextetting order stemming from the frustrations in the phase coherence of a pair-density-wave state, 
whose spatial modulation manifests a vortex-antivortex honeycomb lattice.
The classical ground state configurations are mapped to Baxter's three-coloring model, revealing a macroscopic degeneracy accompanied by extensive entropy. 
The phase coherence problem intertwines the U(1) phases and the vorticity variables.
While the resultant color and phase fluctuations suppress the pair-density-wave order, they maintain the sextetting order above the superconducting transition temperature ($T_{\text{c}}$). 
The $1/3$-fractional vortex emerges as the fundamental topological defect in the sextetting order.
This novel mechanism of frustrated superconductivity provides an alternative explanation for the experimental observed fractional oscillations in CsV$_3$Sb$_5$.
\end{abstract}
%\pacs{}
\maketitle

\section{Introduction}
The multi-fermion ordering has received considerable attention in a wide range of research fields in modern physics.
Inside a heavy nucleus, the alpha-particle-like quartetting
instability involving two protons and two neutrons competes with the deuteron-like instability of pairing between one proton and 
one neutron \cite{Ropke1998}.
Its many-body version has been investigated in the
1D spin-$\frac{3}{2}$ ultra-cold fermion system via bosonization, revealing the Ising-dual relation
between these two instabilities \cite{Wu2005}.
In the context of superconductivity (SC) in electron systems, 
the multi-fermion ordering, quartetting or sextetting, can emerge as a vestigial secondary order of the charge-$2e$ Cooper pairing SC above the superconducting transition temperature $T_{\text{c}}$, often termed as the so-called 
charge-$4e$, or, charge-$6e$ SC states \cite{Korshunov1985,Kivelson1990,Aligia2005,Agterberg2008,Berg2009,Agterberg2011,Wen2009,Herland2010,You2012,Jiang2017, Zeng2021,Fernandes2021,Jian2021,volovik2023fermionic}.

The quartetting or sextetting states can appear 
as a consequence of melting the pair-density-wave 
(PDW) state, i.e., the superconducting pairing
orders $\Delta_{\mathbf{Q}}$ carrying finite momenta
\cite{Agterberg2008,Berg2009, Agterberg2011}. 
A PDW state can exhibit multiple 
incommensurate wave vectors $\mathbf{Q}$, and
the incommensurabilities render the relative phases of the translational degree of freedom gapless.
There exist topological defects related to the translational
phase, and the system undergoes a Kosterlitz-Thouless (KT) type transition \cite{kosterlitz1973ordering} as temperature increases.
The PDW superconducting order disappears above a critical temperature $T_{\mathrm{c}}$ due to the strong fluctuations in the relative phase channels, 
nevertheless, the overall phase remains ordered, 
leading to the quartetting or sextetting state.

Here we propose an alternative mechanism based on the frustrations of the superconducting phase coherence, 
which naturally leads to the sextetting order.
Usually the frustrated systems are characterized by
the macroscopic degeneracy of 
ground-state configurations at the classic level \cite{moessner2006geometrical,balents2010spin}. 
The three-coloring model on the honeycomb lattice \cite{baxter1970} 
is a celebrated example, which applies to 
various frustrated
systems including classical antiferromagnets \cite{huse1992classical}, Josephson junction arrays\cite{DHLee2004,castelnovo2004dynamical}, 
orbital-active materials\cite{wu2007flat,wu2008orbital,wu2008p}, 
and other systems \cite{chern2013frustrated,chakraborty2002topological,You2012}.
Each bond of the honeycomb lattice is painted by one of the three colors $\mathbf{R} (\text{red}),\mathbf{G} (\text{green}),\mathbf{B} (\text{blue})$ 
under the so-called ``color constraint" that three bonds connected to
the same vertex are pained by different colors.
The classical ground-state configurations exhibit a macroscopic degeneracy, leading to an extensive entropy of $0.38k_{\text{B}}$ per hexagon\cite{baxter1970}.
The strong frustrations kill the order of the single color
variable, but
a combined three-color order could 
survive.
This frustration mechanism has also been extended to the 
four-coloring model on the three-dimensional diamond lattice \cite{Chern2014},  
which exhibits the dipolar correlation
characterized by a cubic power decay $1/R^3$.

In this article, we investigate the sextetting order in the superconductivity via frustrations to
the PDW order.
The corresponding superconducting phase coherence problem is mapped to the three-coloring model, exhibiting macroscopic degeneracy arising from color configurations.
The low-energy degrees of freedom encompass the local pairing-phases coupled to the vorticity variables.
Monte-Carlo simulations are performed to obtain
the phase diagram, showing the competition
between the Cooper pairing and sextetting order.
The entropy associated with the color configurations dominates, leading to the sextetting state at a higher temperature before the system enters the completely disordered regime.
Moreover, the sextetting order manifests $1/3$-fractional vortex, providing an alternative mechanism to the fractional flux oscillations.

The proposed mechanism of sextetting order through frustrated superconductivity in the PDW state differs from previous scenario \cite{Agterberg2008,Berg2009,Agterberg2011}. 
Previous analysis primarily focused on the generation of fractional vorticities due to the phase winding among the relative phases, which only involve a few superconducting gap functions. 
In this article, frustration of phase coherence problem in the vortex/antivortex lattice is related to the macroscopic number of degrees of freedom from the vortices/antivortices. 
The number of degenerate states at the classic level grows up exponentially as the lattice size increases, which yields extensive entropy.

\section{Vortex-antivortex honeycomb lattice} 
\begin{figure}[t!]
\centering
\includegraphics[width=0.9\linewidth]{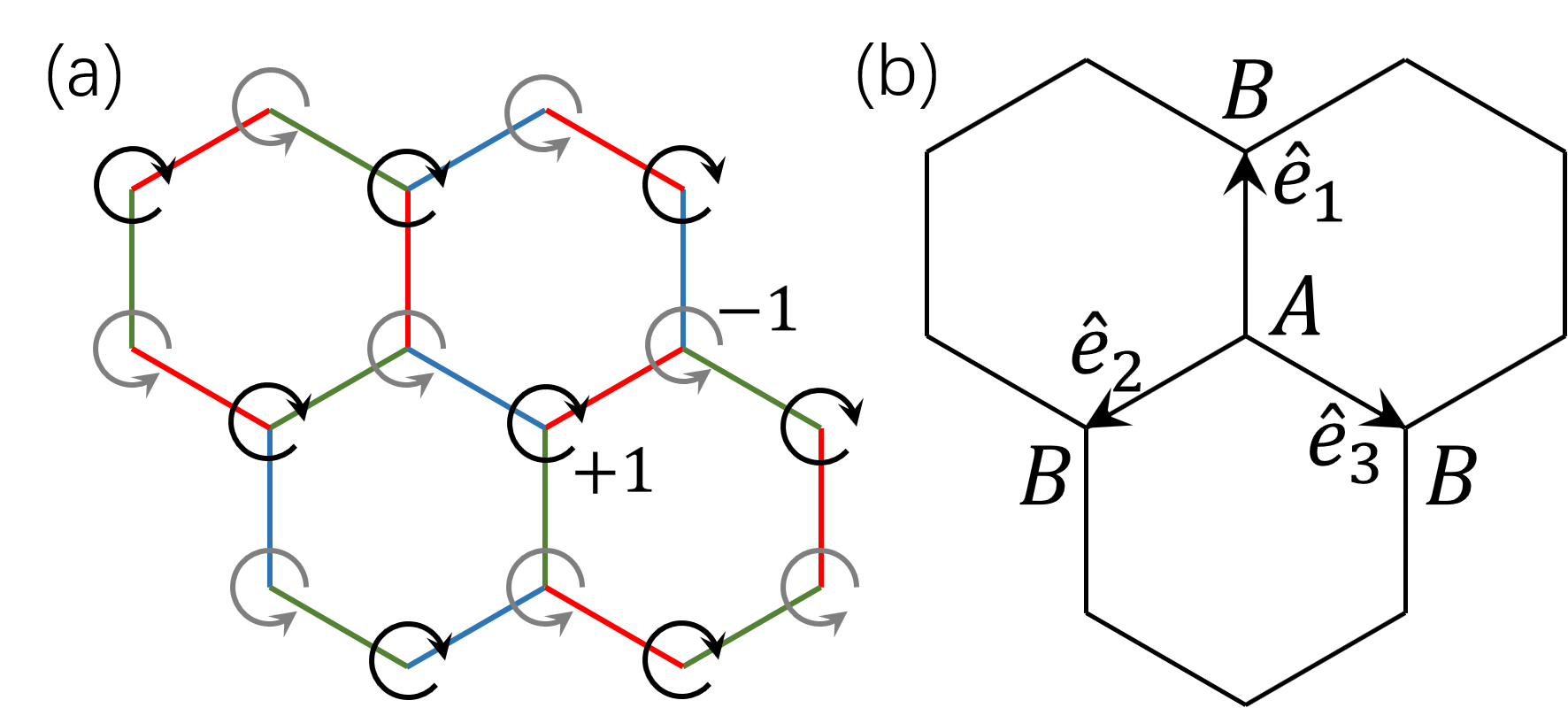}
\caption{(a). A honeycomb lattice of vortices/antivortices.
Here, an AFM-like vortex-antivortex pattern satisfying the three-color constraint is shown.
The superconducting phases along three bond directions can 
take values of $0,\frac{2}{3}\uppi, \frac{4}{3}\uppi$
denoted by three colors $\mathbf{R},\mathbf{G},\mathbf{B}$.
(b). The honeycomb lattice with $A,B$ sublattice.
The three bond directions are denoted as $\hat{e}_{\nu}$ ($\nu=1,2,3$).
}
\label{fig:Figure1}
\end{figure}

\ \ \
Frustrated superconductivity emerges from phase coherence of a vortex-antivortex honeycomb lattice.
The vortex lattice model could be realized through Josephson junction arrays\cite{DHLee2004}, $p$-orbital unconventional superfluidity \cite{wu2009unconventional} or PDW state\cite{Agterberg2011,zhou2022chern}.
The cores of single vortices/anti-vortices are located at the sites
of the honeycomb lattice, 
with phase winding $\pm 2{\uppi}$ around it.
A pair of neighboring single vortices couple with each other through Josephson coupling and inter-vortex coupling.
In the ground state, 
all the vortices and antivortices exhibit phase coherence, 
leading to unconventional superconductivity with simultaneously time-reversal symmetry breaking \cite{wu2009unconventional}.
A typical vortex-antivortex pattern with alternating vortices and anti-vortices is illustrated in Fig.~\ref{fig:Figure1}(a). 
Additionally, the superconducting %phase order 
state induced by a fixed coherent arrangement of vortices and anti-vortices can exhibit spatial modulation, giving rise to the pair-density-wave order.

%------------------------------------------------
%--------------------------------------------------

The vortex-antivortex model can be mapped to the three-coloring problem on 
the honeycomb lattice \cite{DHLee2004,wu2009unconventional}.
(Its generalization to the four-coloring problem in the diamond lattice was
done by Chern and Wu \cite{Chern2014}.)
The superconducting phase winds $\pm 2\uppi$ around a site $j$ of the 
honeycomb lattice, depending on whether a vortex or antivortex is located at that site.
Each site is connected to its three nearest neighbors and 
the angles between two bond directions are all
120$^\circ$. 
Hence, to minimize the phase difference across each bond, without loss
of generality, we can assign three colors $\mathbf{R},\mathbf{G},\mathbf{B}$ 
corresponding to the phases $1$, $\mathrm{e}^{i\frac{2}{3}\uppi}$, $\mathrm{e}^{i\frac{4}{3}\uppi}$, respectively,
and paint each bond according to this color constraint. 
The clockwise or anticlockwise ordering of the colors $\mathbf{R},\mathbf{G},\mathbf{B}$ defines the chirality $\tau_j=\pm 1$,
or, the vorticity. 
Under time-reversal transform, the vorticities are flipped. 
If only considering the color frustration effect, the 
charge-$2e$ correlation of the phases exhibits power-law decay, 
$G_{2e}(r)\sim r^{-4/3}$ \cite{huse1992classical}.

To fully describe the phase coherence of the superconducting state, it is essential to account for U(1) phase fluctuations. 
For later convenience, the following convention is employed:
The honeycomb lattice system is divided into two sublattices $A$ and $B$, as depicted in Fig.~\ref{fig:Figure1}(b).
Considering a site $j$ belonging to sublattice $A$, it emits three bonds denoted as $\hat e_\nu$ whose azimuthal angles are 90$^\circ$, 210$^\circ$, 330$^\circ$ for bond index $\nu=1,2,3$, respectively.
The superconducting phase along the $\hat e_1$ direction is
denoted as $\theta_j$.
On the other hand, for a site $k$ belonging to 
sublattice $B$, its superconducting phase along
$-\hat e_1$ direction is set as $\theta_k$.
Then around any site $j$, the superconducting phase
along the bond $\nu$ is represented as
$\theta_j + \tau_j \phi_\nu$, where $\phi_\nu=0,
\frac{2}{3}\uppi, \frac{4}{3}\uppi$ for $\nu=1,2,3$,
respectively.
Then the effective Hamiltonian reads \cite{wu2009unconventional},
\bea
\begin{aligned}
&H= -J_1 \sum_{j\in A,\nu}
\cos\big[\theta_j-\theta_k + (\tau_{j}-\tau_{k}) 
\phi_{\nu} \big]	\\
-&J_2 \sum_{j\in A,\nu}
\cos\big[\theta_j-\theta_k + (\tau_{j}-\tau_{k}) 
\phi_{\nu}
+\frac{\uppi}{2} (\tau_j+\tau_k) \big],
\end{aligned}
\label{eq:chiralitymodel}
\eea
where the summation over $j$ is only carried on sublattice $A$, and $k$ belongs to sublattice $B$ along the bond direction
of $\hat e_\nu$.

The $J_1$-term represents the nearest-neighbor Josephson coupling,
whose dependence on the bond geometry can be viewed
as a vector potential $A_{jk}=(\tau_j-\tau_k)\phi_\nu$.
$J_1$-term characterizes the phase coherence along the bond direction, enforcing the ``color constraint" within the classical ground state.
The $J_2$-term coupling arises from the non-trivial winding numbers, or angular momenta, of the vortices, similar to the phase coherence problem encountered in orbital physics \cite{wu2009unconventional}.
This term reflects the phase coherence along the bisection line of the bond between two nearest-neighbor vortices.
Typically, in realistic systems, $J_2$ is much smaller than $J_1$-term.
A slight $J_2$-term lifts the degeneracy among color configuration and stabilizes the vortex configuration, fixing it into an antiferromagnetic (AFM)-like 
staggered vortex-antivortex pattern. 
In this pattern, adjacent neighbors display opposing vorticities, as illustrated in Fig.~\ref{fig:Figure1}(a).

\begin{figure}[t!]
\centering
\includegraphics[width=0.8\linewidth]{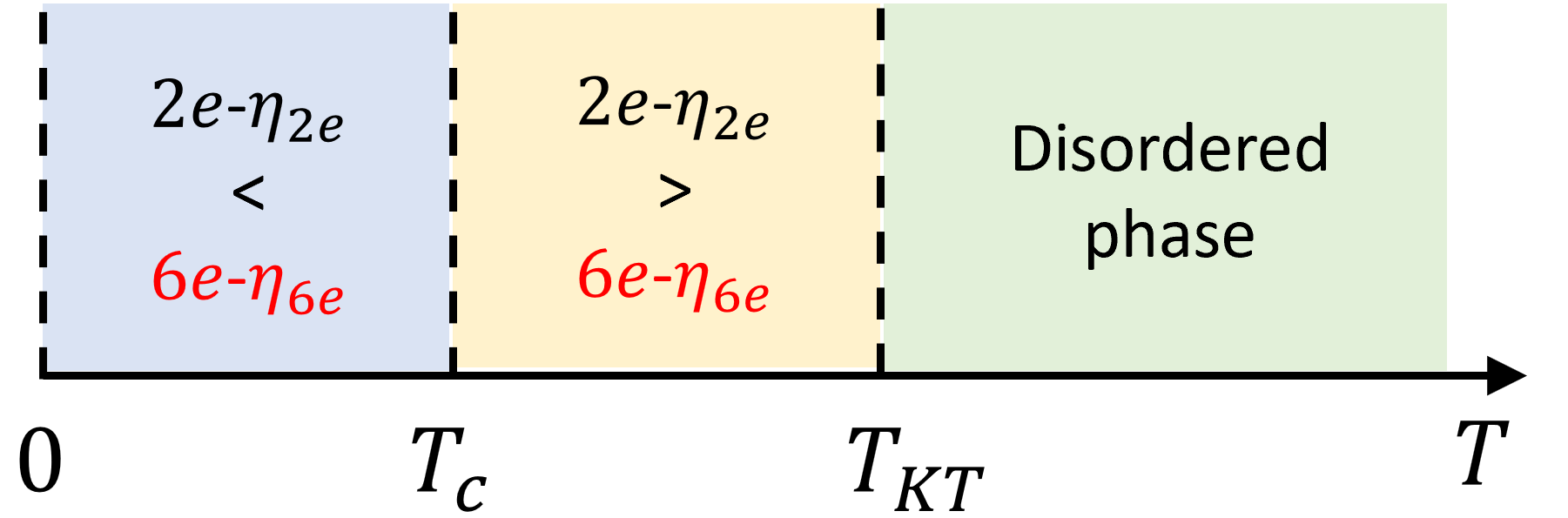}
\caption{A schematic phase diagram of the effective model in Eq.~(\ref{eq:chiralitymodel}).
At low temperatures ($T<T_{\mathrm{c}}$), $2e$ quasi-long-range order (QLRO) dominates ($\eta_{2e}<\eta_{6e}$), while at intermediate temperatures ($T_{\mathrm{c}}<T<T_{KT}$), $6e$ QLRO takes precedence.
Above the KT temperature $T_{\text{KT}}$, the system is in the disordered phase.
}
\label{fig:Figure2}
\end{figure}

If the temperature $T$ is lower than the KT transition temperature
$T_{KT}$ of the U(1) phase $\theta_i$, but is still high enough such that the %long-range 
inter-vortex interactions $J_2$-term
can be neglected, then
the entropy of the color configurations dominates.
As a result, there is no long-range pairing order, but the
cube of the pairing order is long-range ordered
since color variables $1,\omega,\omega^2$ are cubic roots of $1$ with
$\omega=\mathrm{e}^{i\frac{2}{3}\uppi}$.
This manifests as the sextetting superconductivity.
As the temperature is lowered enough, the inter-vortex $J_2$ coupling 
will lift the degeneracy among different vortex configurations.
The ordered PDW configuration emerges, resulting in the transition to the charge-$2e$ superconducting state. 
A schematic phase diagram of the system is summarised as in Fig.~\ref{fig:Figure2}.

%------------------------------------------------
%------------------------------------------------

\section{Numerical simulation}
Monte-Carlo simulations are performed to investigate the Hamiltonian of Eq.~(\ref{eq:chiralitymodel}).
Importantly, the
method of updating
configurations depends on the relevant temperature regime.
Similar situation appears in many frustrated systems, e.g., in the Ref.~\cite{chern2013frustrated}: The acceptance ratio of local
updates becomes vanishingly small at low temperatures which cannot sample the macroscopic degeneracy of the ground states.
The same strategy in Ref.~\cite{chern2013frustrated},
will be followed by
separating the temperature into lower and higher regimes.
Above the KT temperature of the U(1) phase $\theta_i$, $1/3$-fractional vortices proliferate and the U(1) phase coherence is destroyed, which also disorders
the color configurations.
The U(1) phase correlation, or charge-$2e$ correlation, $\langle \mathrm{e}^{i\theta_j} \mathrm{e}^{-i\theta_k}\rangle$ between two sites $j,k$ decays exponentially as the relative distance increases.
The local update of the chiralities and phase variables 
works well in the disordered phase.
As temperature decreases, the vorticity configuration tends to satisfy the color constraint and the local update breaks down\cite{chern2013frustrated}.
Alternatively, we apply the loop update in which the color constraint is maintained in the vorticity configuration in each update.
The U(1) phase variables are updated simultaneously according to the vorticity configuration. (see App.A for the details).

\begin{figure}[t!]
\centering
\includegraphics[width=1.0\linewidth]{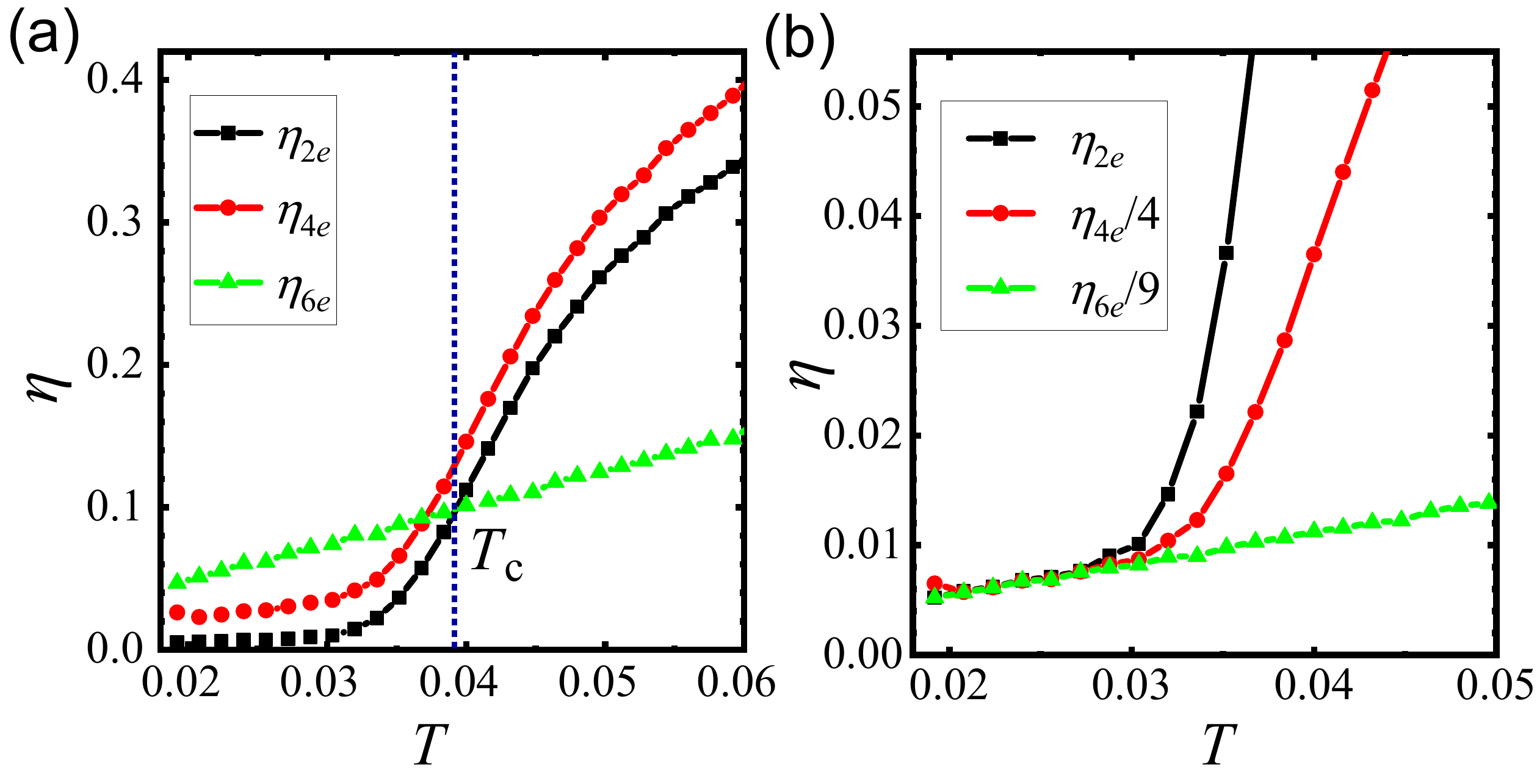}
\caption{(a). Scaling dimensions $\eta$ for $2e$-, $4e$- and $6e$-orders evaluated from the model (\ref{eq:chiralitymodel}) at $J_1=1$ and $J_2/J_1=0.01$ in the temperature regime well below $T_{KT}$.
At temperatures below $T_{\mathrm{c}}$, charge-$2e$ correlation exhibits stronger strength compared to that of charge-$6e$ ($\eta_{2e}<\eta_{6e}$).
The effect of color frustration emerges
above $T_{\mathrm{c}}$ but much lower than $T_{KT}$,
in which charge-$6e$ correlation dominates. 
(b). $\eta_{2e}$, $\eta_{4e}/4$ and $\eta_{6e}/9$ are presented for comparison.
At low enough temperature ($T\lesssim 0.03$), these quantities converge to nearly identical values, i.e., color configuration becomes ordered.}
\label{fig:Figure3}
\end{figure}

The simulations to the scaling dimensions for the superconducting phase correlations in the low temperature regime are summarized in Fig.~\ref{fig:Figure3}(a).
It shows two competing 
phases: the lower temperature charge-$2e$ superconductivity, and the higher temperature charge-$6e$ dominated phase. 
Correlations of both orders are characterized by the quasi-long-range ordering
(QLRO) of the relevant phase correlations defined as
\bea
\begin{aligned}
\langle \mathrm{e}^{i\theta_{j}} \mathrm{e}^{-i\theta_{k}} \rangle \sim r^{-\eta_{2e}}, \quad
\langle \mathrm{e}^{3i\theta_{j}} \mathrm{e}^{-3i\theta_{k}} 
\rangle\sim r^{-\eta_{6e}},
\end{aligned}
\eea
respectively, where $r$ is the distance between the two sites $j,k$.
Here, $\eta_{2e}$ and $\eta_{6e}$ are exponents for the charge-$2e$ and $6e$ correlations, representing the scaling dimensions of the orders.
For comparison, the %critical 
exponent $\eta_{4e}$ of the $4e$-correlation, $\langle \mathrm{e}^{2i\theta_{j}} \mathrm{e}^{-2i\theta_{k}} \rangle \sim r^{-\eta_{4e}}$, is also deduced from the numerical estimation.
To suppress finite size effect and obtain the power-decay exponents more precisely, the numerical simulation is performed at different lattice sizes $L$ with the power-law fitting through the largest distance at $r=(L/2,L/2)$. (see App. B for the technical details).

%------------------------------------------------
%-----------------------------------------

The three %critical 
exponents fit well with the following relations,
\bea
\eta_{2e}=\eta_{\mathrm{c}}+\eta_{6e}/9, \quad
\eta_{4e}=\eta_{\mathrm{c}}+4\eta_{6e}/9. 
\eea
Here, $\eta_{\mathrm{c}}$ represents the contribution from color frustration in the vorticity correlation, which does not contribute to the charge-$6e$ correlation.
At low enough temperatures, the vortex configuration is fixed with $\eta_{\mathrm{c}}=0$ and the exponents arise from the QLRO of the U(1) phase variables, exhibiting $\eta_{2e}=\eta_{4e}/4=\eta_{6e}/9$, as depicted in Fig.~\ref{fig:Figure3} (b). 
It is indicated that charge-$4e$ is never the leading correlation.
The most dominant correlation is either charge-$2e$ at low temperatures, or charge-$6e$ at high temperatures.

The transition from charge-$2e$ to charge-$6e$ is driven by the competition between the energy consideration and entropy contribution.
At lower temperatures, energetic minimization stemming from the combination of $J_1$ and $J_2$ terms favors the AFM-like staggered ordering of the vortex/anti-vortex configuration.
The low-temperature regime is characterized by the dominance of charge-$2e$ behavior ($\eta_{2e}<\eta_{6e}$).
As temperature increases, color frustration begins to manifest.
Configurations satisfying the three-coloring constraint minimize $J_1$-term ($J_2$ is much smaller than $J_1$). 
The number of such kind of configurations scales exponentially as increasing the system size, exhibiting an extensive entropy.
The scaling dimensions of $2e$ and $6e$ correlations intersect at a transition temperature $T_{\mathrm{c}}$.
Above $T_{c}$, but still below the KT temperature, the entropy associated with color configurations takes precedence over the energy cost arising from the $J_2$-term and the charge-$2e$ correlation is suppressed.
The charge-$6e$ behavior dominates over the charge-$2e$ correlation, as evidenced by the relationship $\eta_{6e}<\eta_{2e}$.
Moreover, the estimated color correlation $\eta_{\mathrm{c}}$ is much smaller than $4/3$ for the pure three-color model\cite{huse1992classical}.
This is because the degeneracy among different color configurations is broken by the $J_2$ term.

The behaviors of the scaling dimensions related to $2e$, $4e$ and $6e$ order under temperature for smaller $J_2$ exhibit qualitatively consistent behavior as in Fig.~\ref{fig:Figure3}, see simulations for $J_2/J_1=0.015,0.02$ in App. C.
As $J_2$-term increases, the energy cost lifting the degeneracy of the color configurations increases, leading to an enhancement of $T_{\mathrm{c}}$.
When the $J_2$-term becomes large enough, the energetic consideration could always outweigh the entropy contribution, resulting in the absence of charge-$6e$ QLRO, as depicted in the simulation result for $J_2/J_1=0.03$ in App. C.
In reality, $J_2$ coupling along the bisection line of the bond is always exists, while its amplitude is typically much smaller than $J_1$ term, aligning with the scenario depicted in Fig.~\ref{fig:Figure2}.

Above the KT temperature $T_{KT}$ of the U(1) phase variables $\theta_i$, both charge-$2e$ and $6e$ correlations exhibit exponential decay, leading the system into a disordered phase.
%The color configuration also becomes disordered and a local update proves sufficient for estimating correlation behavior above $T_{KT}$.
The numerical simulation confirms the exponentially decay behavior of the correlations in this regime.
However, due to limitations in Monte Carlo simulation, as mentioned earlier, precise determination of the KT temperature is not feasible. 
As the temperature decreases from high to the critical temperature $T_{KT}$, the effectiveness of local updates gradually diminishes, reflected in rapidly increasing acceptance rates and computation times.
The temperature at which the local updates become ineffective is approximately around $T_0\approx 0.13$, establishing an upper bound for $T_{KT}$.
This kind of numerical difficulty is common in frustrated system
\cite{chern2013frustrated}.
More advanced method is necessary for completely simulating the problem, which is left for further works.

The preceding analysis focuses on the inherently two-dimensional system, where charge-$2e$ and charge-$6e$ correlations exhibit only quasi-long-range order. 
In the real materials, it is natural that the system comprises a layered 2D lattice, thereby forming a quasi-two-dimensional model.
The inter-layer couplings between the 2D layers enhance the QLRO, stabilizing a true long-range order of either the charge-$2e$ type (for $\eta_{2e}<\eta_{6e}$ regime), or, charge-$6e$ type (for $\eta_{6e}<\eta_{2e}$ regime) at low temperatures.

The $\frac{1}{3}$-fractional flux can emerge above the $T_{\mathrm{c}}$ of 
the PDW order due to the vortex configuration fluctuations \cite{DHLee2004,wu2009unconventional}.
Around each hexagonal vortex/antivortex plaquette $h$, its total vorticity is defined as $\Phi_h=\frac{1}{3}\sum_{i\in h}\tau_i$.
For a configuration satisfying the color constraint, the vorticity around a plaquette takes the values of $0, \pm 2$.
A fundamental topological defect exhibits a domain wall
which is a string of bonds with mismatched color variables.
The starting and the ending points
of a domain wall are located in
a plaquette as the core of 
a $\frac{1}{3}$-vortex, or, a 
$\frac{1}{3}$-antivortex,
respectively.
Around the vortex cores, the superconducting phase winding
equals to $\pm \frac{2}{3}\uppi$,
respectively.
An external magnetic flux at 
1/3 of the fundamental flux generates a $\frac{1}{3}$-vortex such that
the superconducting phase across each bond remains matched.
Consequently, the free energy will show periodic modulation under the flux change of
\bea
\Phi_{6e}=\frac{1}{3}\Phi_{2e}=\frac{hc}{6e},
\eea
exhibiting the charge-$6e$ QLRO and $1/3$-fractional flux.

\section{PDW states}
PDW states exhibiting spatial modulation of the pairing order have been proposed in the FFLO states \cite{fulde1964FF,larkin1965LO}.
The vortex-antivortex model provides a platform for realizing the PDW state as its ground state. 
In the hexagonal crystal system, PDW state can manifest $6$ commensurate wave vectors denoted as
$\pm \mathbf{Q}_i$ ($i=1,2,3$).
The spatial distribution of PDW order parameter is
represented as \cite{Agterberg2011},
\bea
\Delta(\mathbf{r})=\sum_i \left(\Delta_{\mathbf{Q}_i} \mathrm{e}^{i\mathbf{Q}_i\cdot\bm{r}}
+\Delta_{-\mathbf{Q}_i} 
\mathrm{e}^{-i\mathbf{Q}_i\cdot\bm{r}} \right),
\label{ChiralPDW}
\eea
where the gap functions $\Delta_{\pm\mathbf{Q}_i}$ ($i=1,2,3$) carry nonzero mass-of-center
momentum $\pm\mathbf{Q}_i$, respectively.
The ground state configuration can be determined by the Ginzburg-Landau free energy \cite{Agterberg2011}, presenting several possibilities depending on the microscopic details of the relevant model.

\begin{figure}[t!]
\centering
\includegraphics[width=0.9\linewidth]{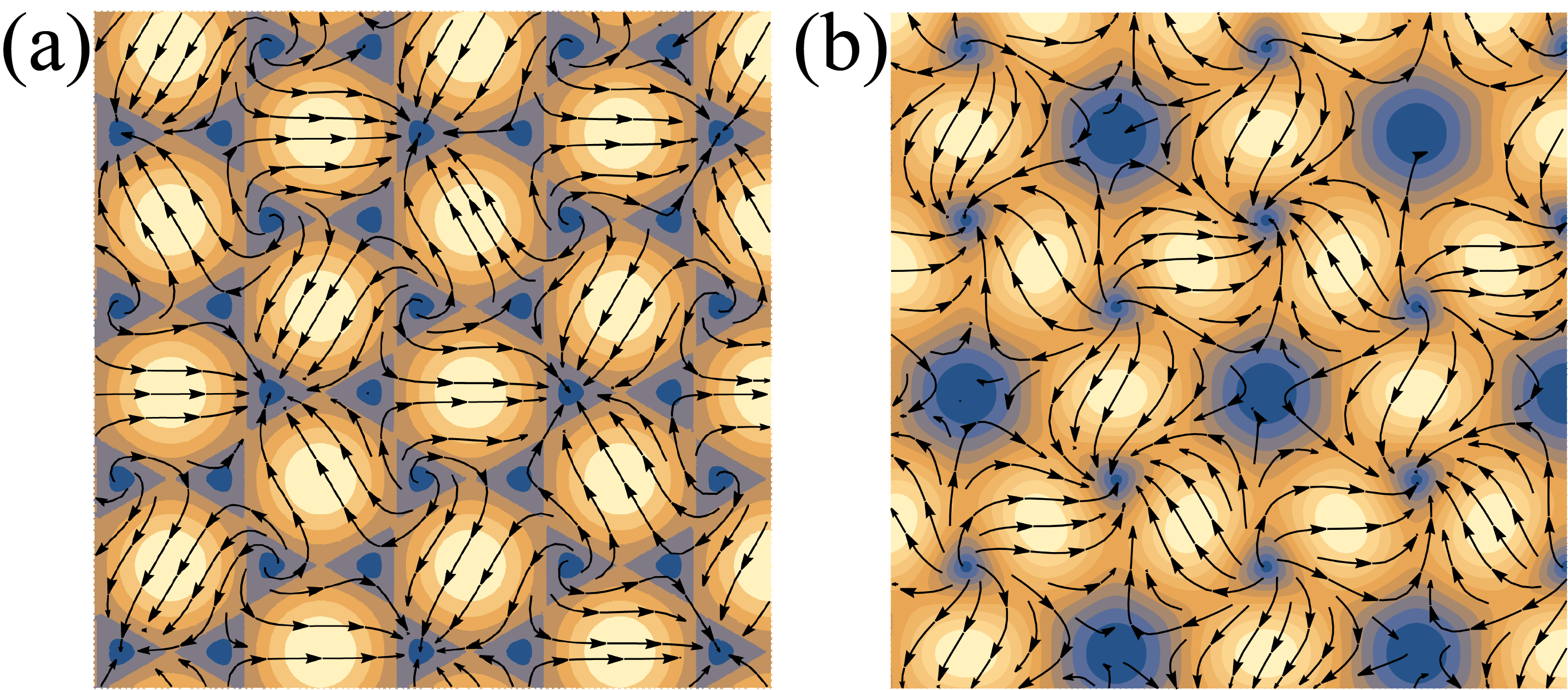}
\caption{The vortex-antivortex lattice pattern of the PDW state for the (a) 3Q state and (b) 6Q state. In the 3Q state, single vortices and antivortices form a honeycomb lattice.
In the 6Q state, single vortices (or antivortices) form a honeycomb lattice with double antivortices (or vortices) in the centers of the hexagons.}
\label{fig:Figure4}
\end{figure}

For the previously considered AFM-like vortex-antivortex pattern, it corresponds to an inversion symmetric breaking PDW state with only three wavevectors $\mathbf{Q}_1,\mathbf{Q}_2$ and $\mathbf{Q}_3$.
The superconducting order parameter is given by
\bea
\Delta_{3\mathbf{Q}}(\bm{r})
=\Delta_{3\mathbf{Q}} \big(
\mathrm{e}^{i\mathbf{Q}_1\cdot\bm{r}}
+ \mathrm{e}^{i\theta_2+i\mathbf{Q}_2\cdot\bm{r}}
+ \mathrm{e}^{i\theta_3+i\mathbf{Q}_3\cdot\bm{r}} \big),
\eea
with relative phases $\theta_2,\theta_3$ between the three PDW components.
This 3Q state forms a honeycomb lattice with alternating vortices/antivortices as illustrated in Fig.~\ref{fig:Figure4}(a) \cite{Agterberg2011}.
On the other hand, when inversion symmetry is preserved, the 6Q PDW order parameter exhibits all the six momenta and a typical chiral PDW pattern is depicted in Fig.~\ref{fig:Figure4}(b) \cite{Agterberg2011,zhou2022chern}, where
a double vortex/anti-vortex appears in plaquette
centers. 
The 3Q-state with the AFM-like vortex/antivortex lattice pattern is energetically more favorable than the 6Q-state.
Notice that previous mechanism of frustrated superconductivity could similarly occur in the 6Q-state, where the honeycomb lattice is formed from single vortices or antivortices.

%------------------------
%--------------------------

In the STM experimental detection of the PDW state, the local density of state (LDOS) exhibits extra spatial modulations resulting from the PDW state
\cite{hamidian2016detection,dai2018pair,wang2018pair,edkins2019magnetic,du2020imaging,liu2021discovery,Chen2021,gu2023detection}.
The PDW often coexists with the uniform SC $\Delta_0$, then strong peaks shown in the Fourier transform of LDOS correspond to the PDW wavevectors directly.
The uniform SC component couples to a particular PDW component $\Delta_{\mathbf{Q}}$, contributing to a spatial modulation component $N_{\mathbf{Q}}(\bm{r})$ in the LDOS to the lowest order,
\bea
N_{\mathbf{Q}}(\bm{r})
\propto \Delta_0^*\Delta_{\mathbf{Q}} \mathrm{e}^{i\mathbf{Q}\cdot\bm{r}}
+\Delta_0\Delta_{\mathbf{Q}}^* \mathrm{e}^{-i\mathbf{Q}\cdot\bm{r}}.
\label{eq:LDOSsingleQ}
\eea
The LDOS measurement from a STM experiment can not distinguish the 3Q PDW state from the 6Q one up to this order, since both of them exhibit six peaks in the Fourier spectrum of LDOS.

The difference between the LDOS patterns of the 3Q and 6Q states arises from the couplings among PDW components.
A pair of two PDW components contribute to the LDOS,
\bea
N_{\mathbf{Q}_i-\mathbf{Q}_j}(\bm{r})
\propto \Delta_{\mathbf{Q}_i}^* \Delta_{\mathbf{Q}_j} \mathrm{e}^{i(\mathbf{Q}_i-\mathbf{Q}_j)\cdot\bm{r}}
+\text{h.c.}
\eea
For the 3Q state, the LDOS pattern exhibits peaks corresponding to the wavevector $\pm(\mathbf{Q}_i-\mathbf{Q}_j)$ with $i\neq j=1,2,3$.
For the 6Q state, several extra peaks $\pm 2\mathbf{Q}_i$ ($i=1,2,3$) different from the 3Q state appear.
Such differences could be helpful to distinguish these two PDW states.

\section{Discussions} 
Recently, superconductivity of the series of Kagome materials 
AV$_3$Sb$_5$ (A=K,Rb,Cs) 
has been observed \cite{Ortiz1,Ortiz2,Ortiz3}.
A $\frac{4}{3}\times\frac{4}{3}$ bidirectional spatial modulation of the 
superconducting gap function is observed \cite{Chen2021} in the 
low-temperature superconducting state on the top of a 
$2\times2$ charge-density-wave (CDW) normal-state background \cite{wang2021electronic,Jiang2021,Liang2021,zhou2022chern,song2022orbital}. 
This superconducting state is suggested to be a chiral PDW state which breaks the time-reversal symmetry spontaneously \cite{Chen2021,zhou2022chern}.
However, as described previously, it may not be easy to distinguish the 3Q-PDW and 6Q-PDW orders when only six peaks are clearly seen in the experiment \cite{Chen2021}.

A particularly interesting experimental progress on the CsV$_3$Sb$_5$ superconductor is the detection of the $hc/(6e)$ magneto-resistance oscillation period in the Little-Parks experiment above $T_{\mathrm{c}}$, one 
third of the conventional Little-Parks oscillation \cite{Ge_J2022}.
The phenomena of unconventional fractional oscillations in superconductor have aroused considerable interests \cite{yerin2017anomalous,zhou2022chern,Huxiao2022,Patrick2022,yu2022non,jin2022interplay}.
The above mechanism to the charge-$6e$ state based on the frustrations of the PDW order may be applied to this fractional $hc/(6e)$ oscillation.
The experiment set up is a thin CsV$_3$Sb$_5$ flake with a hole forming a mesoscopic ring. 
Below the $T_{c}$ of the PDW order, an external magnetic flux going 
through the hole generates an integer vorticity, accounting for the conventional ${hc}/{2e}$ oscillation.
As increasing temperature, frustrations suppress the charge-$2e$ PDW
order but favor the charge-$6e$ order.
An external magnetic flux of $n$ times of ${hc}/{6e}$ can generate 
the $n/3$ fractional vorticity, with the core pinned inside the hole, 
which gives rise to the ${hc}/{6e}$ type quantum oscillations.

The above idea of frustrated superfluidity can also be extended to three dimensions. 
The three-coloring model is generalized to the 4-coloring model defined on a diamond lattice, 
and the color constraint is updated as that all the four bonds connected to a lattice site should be painted by 
the $\textbf{R}$, $\textbf{G}$, $\textbf{B}$, $\textbf{Y}$ colors without repetition \cite{Chern2014}. 
The $\textbf{R}$, $\textbf{G}$, $\textbf{B}$, $\textbf{Y}$ colors represent the quartic unit roots $1, i, -1, -i$ respectively.
The frustration therein could favor a charge-$8e$ state.

It is worth noting that the phenomenon of quantum quartetting, or charge-$4e$, order has been investigated in the previous study within the context of one-dimensional spin-$\frac{3}{2}$ fermionic system \cite{Wu2005}.
A ground state featuring charge-$4e$ order driven by quantum fluctuations was explicitly identified in certain parameter regimes.
It competes with the charge-$2e$ pairing as tuning interaction parameters.
Since the charge-$4e$ ordering can be viewed as pairing of pairs, the transition between charge-$2e$ and $4e$ orderings is Ising-like, i.e., these two phases are Ising-dual to each other.
The model proposed in Eq.~(\ref{eq:chiralitymodel}) could also be augmented to the quantum version, for example, by adding the on-site Bose Hubbard interaction.
Typically, “order-from-disorder” mechanism will select the charge-$2e$ ordering state in $(2+1)$-dimensions even in the absence of $J_2$ term.
It would be interesting to study how to further enhance quantum fluctuations to achieve the quantum sextetting order in the future studies.

\section{Conclusions} 
We study the phase frustrations in the vortex-antivortex model on the honeycomb lattice by mapping its
superconducting phase coherence problem to the three coloring model.
The fundamental degrees of freedom are described by the U(1) phases coupled to the discrete vorticity variables. 
The classical ground states, subject to the color constraints, exhibit a macroscopic degeneracy in terms of coloring patterns.
The inclusion of inter-vortex coupling $J_2$ results in a phase coherent ground state pattern that naturally realizes the pair-density-wave state. 
Above the superconducting transition temperature ($T_{\mathrm{c}}$), the sextetting (charge-$6e$) order prevails over the pairing (charge-$2e$) order , persisting until the system enters a disordered regime.
Such behavior is further numerically verified through Monte Carlo simulations.
The fundamental topological defects are $\pm\frac{1}{3}$ vortices,
which leads to the $hc/6e$ flux modulations. 
Our findings establish possible connections to the recent quantum oscillation experiments on the Kagome superconductor CsV$_3$Sb$_5$ \cite{Ge_J2022}.

\section*{Acknowledgements}
{We are grateful to the stimulating discussions with Shaokai Jian and Zheng Yan.
Congjun Wu is supported by the National Natural Science Foundation of
China under the Grant No. 12234016 and No. 12174317.
This work has been supported by the New Cornerstone Science
Foundation. 
Fan Yang is supported by the National Natural Science Foundation of China under the Grant No. 12074031.
Chen Lu is supported by the National Natural Science Foundation of China under the Grant No. 12304180.}

{The authors declare that they have no conflict of interest.}

\setcounter{equation}{0}
\renewcommand{\theequation}{A\arabic{equation}}

\setcounter{figure}{0}
\renewcommand{\thefigure}{A\arabic{figure}}

\section*{Appendix A. Loop update}
In this appendix, we provide a detailed explanation of the combined loop update method employed in the low-temperature Monte Carlo simulations.
At exceedingly low temperatures, the system primarily manifests states within the three-color subspace, adhering to color configurations subjected to color constraints.  
To effectively capture the system's behavior in such conditions, we would confine the simulation to the three-color subspace in this temperature regime.

\begin{figure}[t!]
\centering
\includegraphics[width=0.8\linewidth]{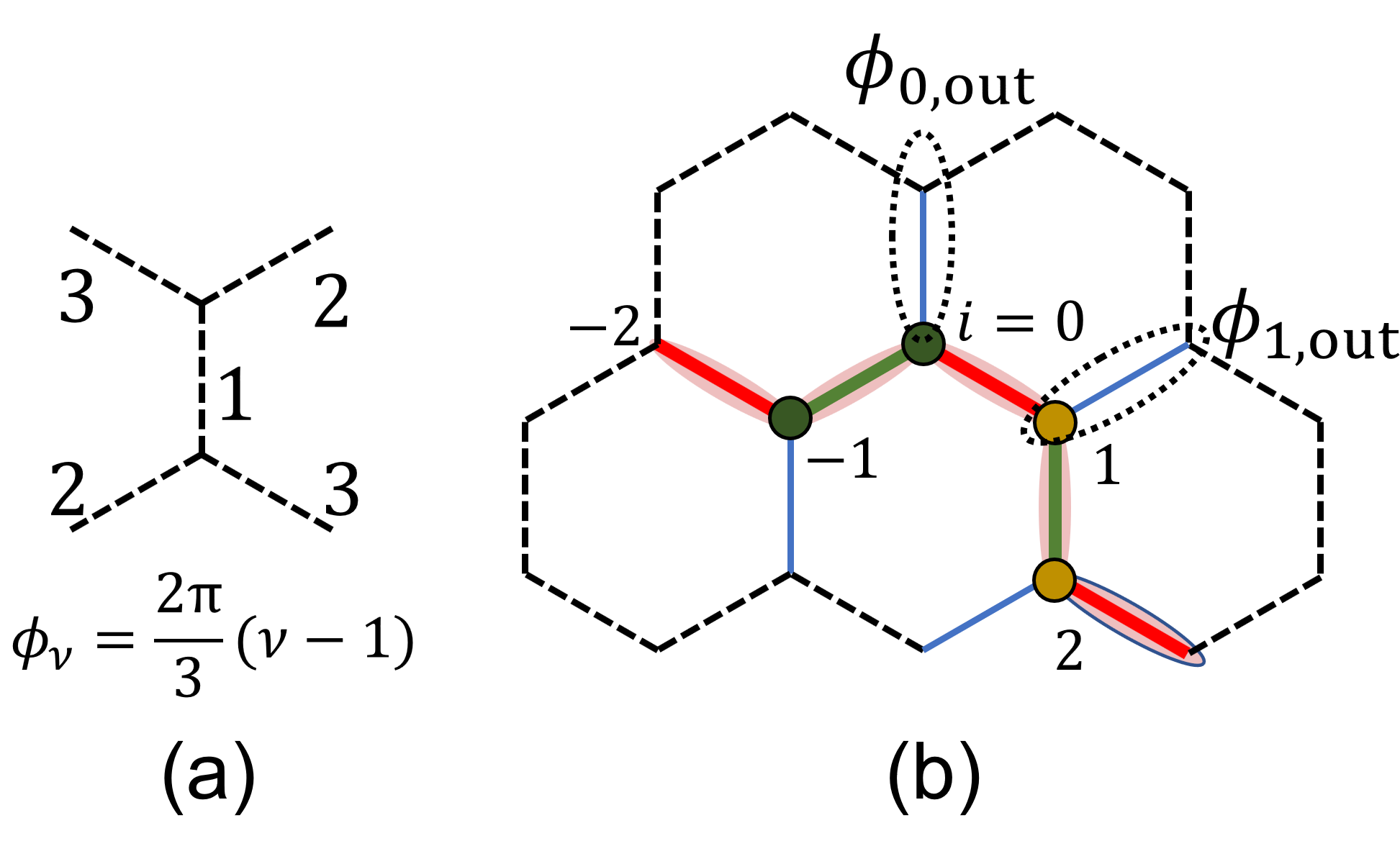}
\caption{(a). The three bonds $\nu=1,2,3$ and the corresponding coupling phase $\phi_\nu=0,\frac{2}{3}\uppi, \frac{4}{3}\uppi$ for them. 
(b). A schematic diagram for a loop with alternating colors $RGRG\cdots$ and the external bond.
The sites $i$ on the loop are labeled by $\cdots,-2,-1,0,1,2,\cdots$.}
\label{fig:Figure_a1}
\end{figure}

For the implementation of the combined loop update method, we initially define loops within the three-color subspace. 
These loops are generated by traversing a path with two alternating colors, such as $RGRG\cdots$, as depicted in Fig.~\ref{fig:Figure_a1}(b). 
Importantly, the interchange of the two colors along the loop does not violate the three-color constraint. 
In the loop update, a specific loop is selected, and colors along its path are interchanged, effectively flipping the chiralities along the loop.

%------------------------------------------
%------------------------------------------
%------------------------------------------
%------------------------------------------

Simultaneous updates of the U(1) phases are also necessary for the sufficiency of the simulation.
During the loop update, the chiralities along the loop $L$ are flipped, and the local U(1) phases $\theta_i (i\in L)$ are altered concurrently. 
It is crucial to note that each site $i\in L$ within the loop possesses one external bond that is not part of the loop, featuring a phase coupling denoted as $\phi_{i,\text{out}}$ at the bond. 
The transformation of $\theta_i (i\in L)$ under the update is determined while keeping the external phase $\varphi_{i,\text{out}}$ fixed,
\begin{eqnarray}
\varphi_{i,\nu=\text{out}}
=\theta_i^{(0)} +\phi_{i,\text{out}} \tau_i^{(0)}
=\theta_i^{(1)} +\phi_{i,\text{out}} \tau_i^{(1)}
\end{eqnarray}
where $(\theta_i^{(0)},\tau_i^{(0)})$ and $(\theta_i^{(1)},\tau_i^{(1)})$ are phase variables and chiralities before and after the loop update.
All the chiralities along the loop are flipped, $\tau_i^{(1)}=-\tau_i^{(0)}$,
and we can find the updated phase $\theta_i^{(1)}$ as
\begin{eqnarray}
\theta_i^{(1)} =\theta_i^{(0)} +2\phi_{i,\text{out}} \tau_i^{(0)}.
\end{eqnarray}
In total, for each loop $L$, we use the following updates for the color configuration,
\begin{eqnarray}
i\in L:\qquad
\begin{cases}
\tau_i^{(0)} &\quad\rightarrow 
\quad\tau_i^{(1)} =-\tau_i^{(0)},	\\
\theta_i^{(0)} &\quad\rightarrow \quad
\theta_i^{(1)} =\theta_i^{(0)} +2\phi_{i,\text{out}} \tau_i^{(0)}.
\end{cases}
\end{eqnarray}

\section*{Appendix B. Finite size effect and fitting of the scaling dimensions}

In the Monte-Carlo simulation, the correlation functions for $2e$, $4e$, and $6e$ are numerical evaluated at different lattice sizes ($L=12, 18, 24, 30, 36, 42, 48$). 
For each size, the correlation functions are calculated based on the following formula,
\begin{eqnarray*}
\begin{aligned}
2e:&\quad F_{2e}(m;L)=\frac{1}{N} \sum_{j} \langle \mathrm{e}^{i \theta_{j}} \mathrm{e}^{-i\theta_{j+m}} \rangle,   \\
4e:&\quad F_{4e}(m;L)=\frac{1}{N}\sum_{j} \langle \mathrm{e}^{i2 \theta_j} \mathrm{e}^{-i2 \theta_{j+m}} \rangle,   \\
6e:&\quad F_{6e}(m;L)=\frac{1}{N}\sum_{j} \langle \mathrm{e}^{i3\theta_j} \mathrm{e}^{-i3\theta_{j+m}} \rangle,
\end{aligned}
\end{eqnarray*}
where $j$ is the lattice site, $m$ is the relative distance and $N=L^2$ is the total number of lattice sites.

%------------------------------------------
%------------------------------------------
\begin{figure}[t!]
\centering
\includegraphics[width=0.9\linewidth]{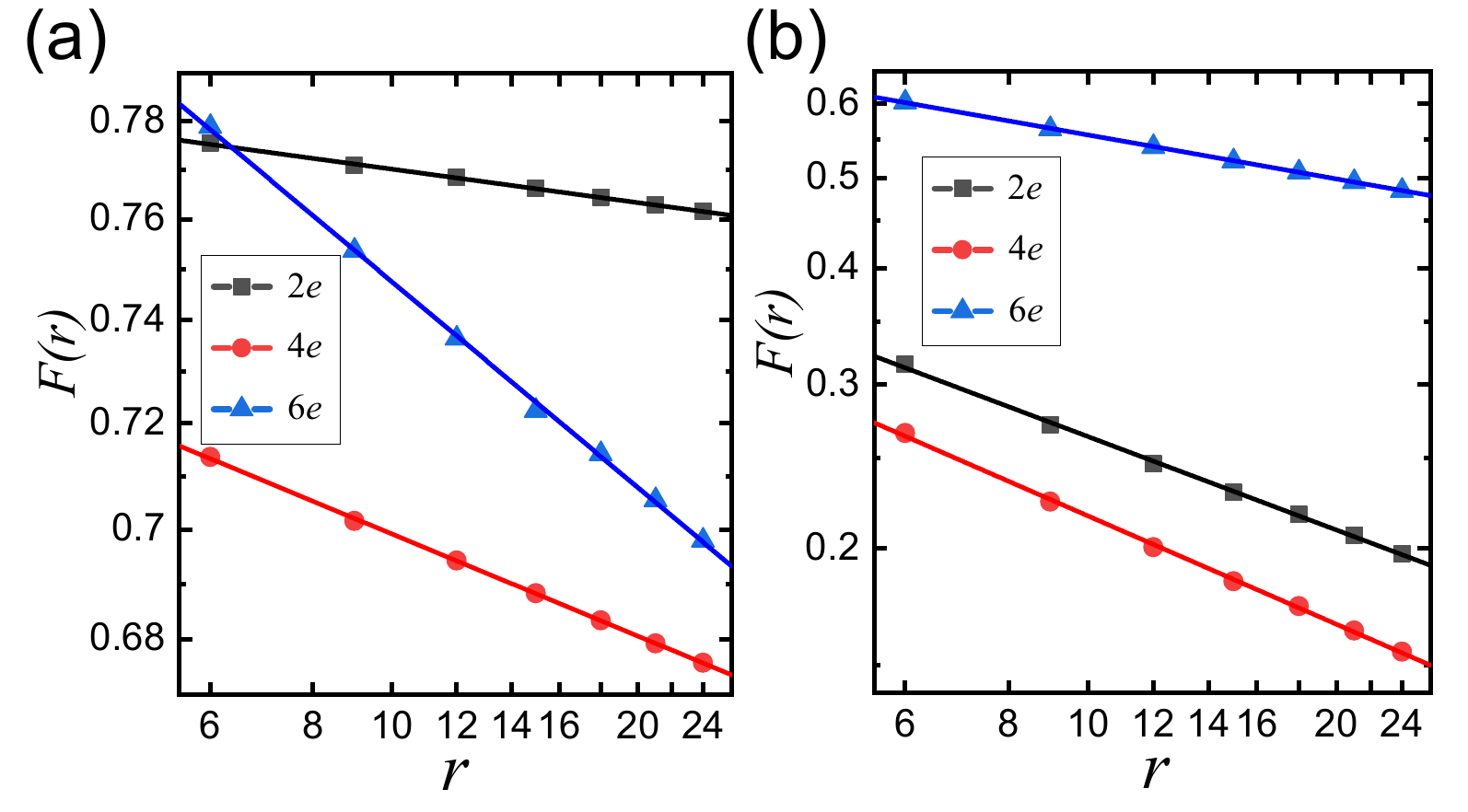}
\caption{Log-log plot of the correlation functions $F(r)$ evaluated for $2e$-, $4e$- and $6e$-orders at (a) $T=0.032$ and (b) $T=0.064$ for $J_1=1$ and $J_2/J_1=0.01$.}
\label{fig:Figure_a2}
\end{figure}
%------------------------------------------
%------------------------------------------

The simulated results $F(m;L)$ for a fixed size $L$ suffer strong finite size effect for larger $m$ closing to $L$ and the directly fitting of exponents would not be good.
Instead, for each lattice size $L$, the value of the correlation functions at the furthest distance, $r=(L/2, L/2)$, is selected, $F(r)\equiv F(r;L)$. 
The correlation exponents are then fitted within $F(r)$ for different lattice sizes. 
This approach effectively reduces finite size effect on the calculations when simulating exponents.
For example, two typical fitting functions $F(r)$ at $T=0.032$ and $T=0.064$ for $J_1=1$ and $J_2/J_1=0.01$ are depicted in Fig.~\ref{fig:Figure_a2}.
At smaller temperature $T=0.032$, $2e$ correlation roughly exhibits larger absolute value and smaller power-law decaying exponent than that of $6e$ correlation.
The system exhibits the $2e$-QLRO state.
On the contrary, at larger temperature $T=0.064$, the above relations reverse and the system exhibits the $6e$-QLRO state.

\section*{Appendix C. Behavior of exponents at difference $J_2$}
\begin{figure}[t!]
\centering
\includegraphics[width=0.95\linewidth]{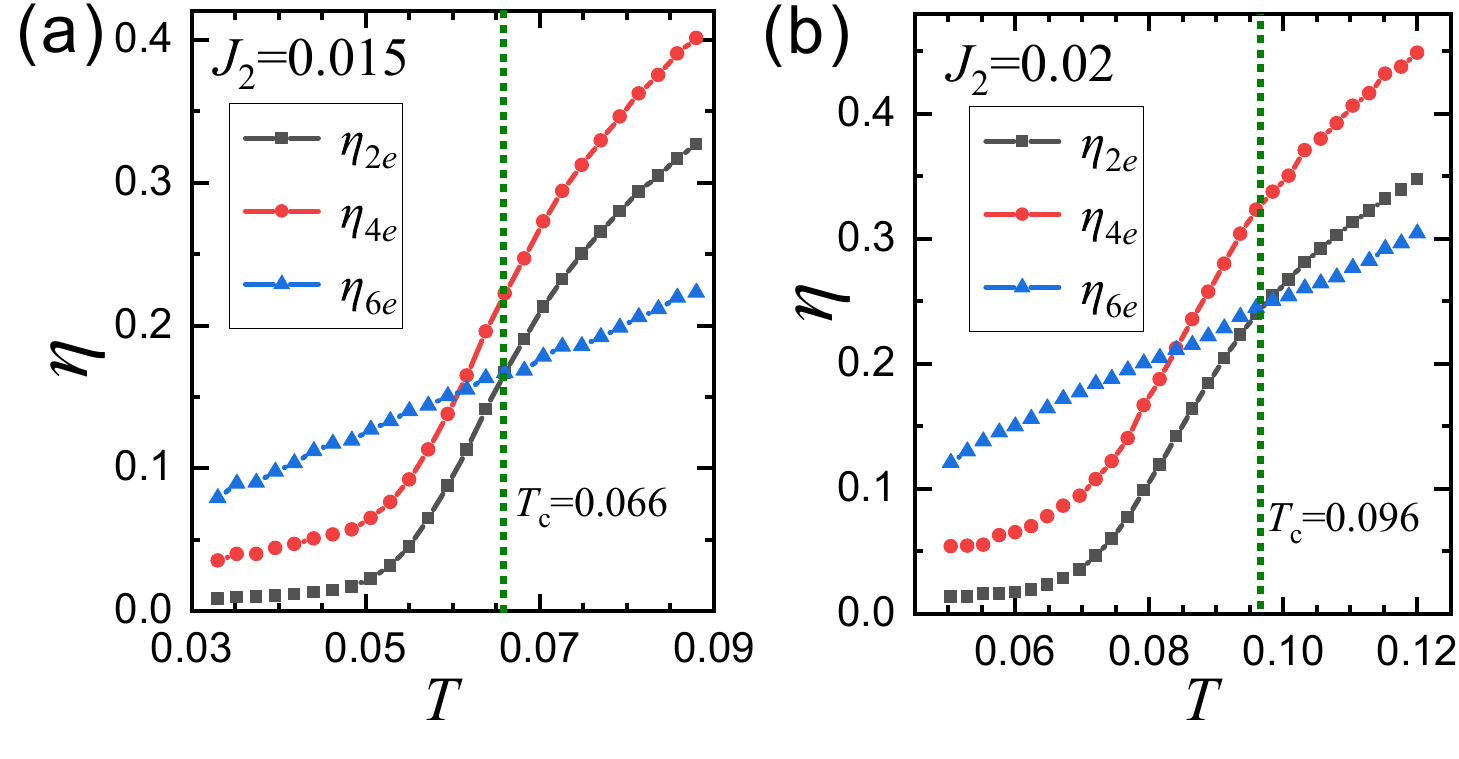}
\caption{ Scaling dimensions $\eta$ for $2e$-, $4e$- and $6e$-orders 
evaluated from the model (\ref{eq:chiralitymodel}) at (a) $J_2/J_1=0.015$ and (b) $J_2/J_1=0.02$ in the temperature regime well below $T_{KT}$.
Here, $J_1=1$ is set as energy scale.}
\label{fig:Figure_a3}
\end{figure}

The simulations to the power-lay decay exponents related to $2e$, $4e$ and $6e$ correlations under temperature for $J_2/J_1=0.015,0.02$ are depicted in Fig.~\ref{fig:Figure_a3}.
These results exhibit qualitatively consistent behavior with $J_2/J_1=0.01$ depicted in Fig.~\ref{fig:Figure2}.
At low temperature below $T_{\mathrm{c}}$, due to the dominance of energy consideration from $J_2$-term, $2e$-QLRO wins over $6e$-QLRO, indicated by the lower scaling dimension $\eta_{2e}<\eta_{6e}$. 
When $T_{\mathrm{c}}<T<T_{KT}$, entropy contribution from color frustration wins over energetic consideration, leading to the $6e$-QLRO with dominant instability, $\eta_{6e}<\eta_{2e}$.
The transition temperature $T_{\mathrm{c}}$ becomes larger as $J_2$ increases, due to the increased energy cost deviation from the staggered vortex/antivortex configuration.

%------------------------------------------
%------------------------------------------
%------------------------------------------
%------------------------------------------

%------------------------------------------
%------------------------------------------
\begin{figure}[t!]
\centering
\includegraphics[width=0.5\linewidth]{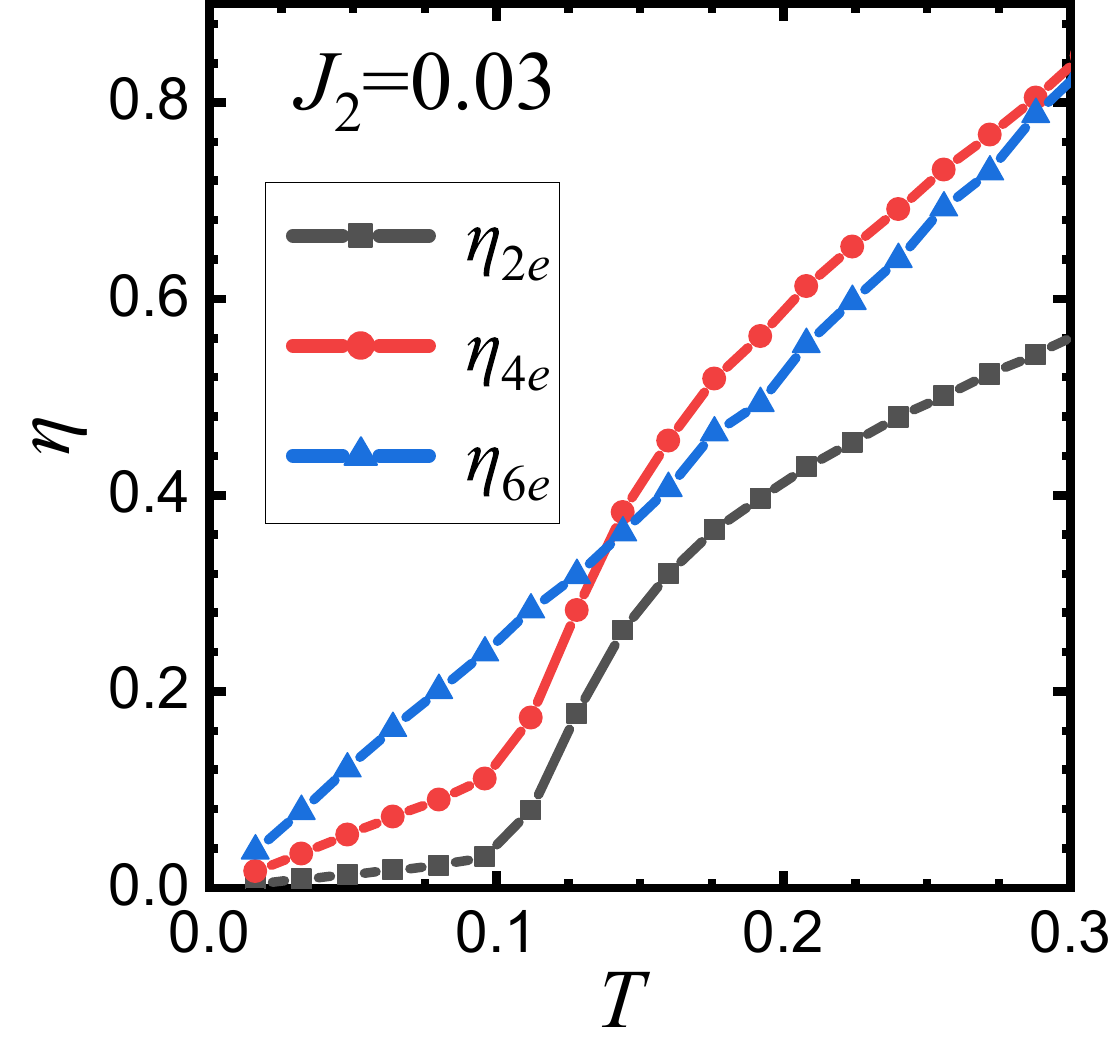}
\caption{Scaling dimensions $\eta$ for $2e$-, $4e$- and $6e$-orders 
evaluated from the model (\ref{eq:chiralitymodel}) at $J_2/J_1=0.03$ in the temperature regime well below $T_{KT}$.}
\label{fig:Figure_a4}
\end{figure}
%------------------------------------------
%------------------------------------------

When $J_2$ surpasses a critical value, the energy cost associated with it could always outweigh the entropy contribution from color frustration across the entire temperature regime, resulting in the absence of intermediate charge-$6e$ QLRO.
For example, for $J_2=0.03J$, the calculated scaling dimensions are depicted in Fig.~\ref{fig:Figure_a4}.
Here, $2e$-QLRO exhibits the dominant instability, indicated by the lowest scaling dimension $\eta_{2e}$ in a broad temperature regime below $T_{KT}$.
Consequently, only the charge-$2e$ state is expected to persist above a certain $J_2$ value at low temperature.

\begin{figure}[t!]
\centering
\includegraphics[width=0.7\linewidth]{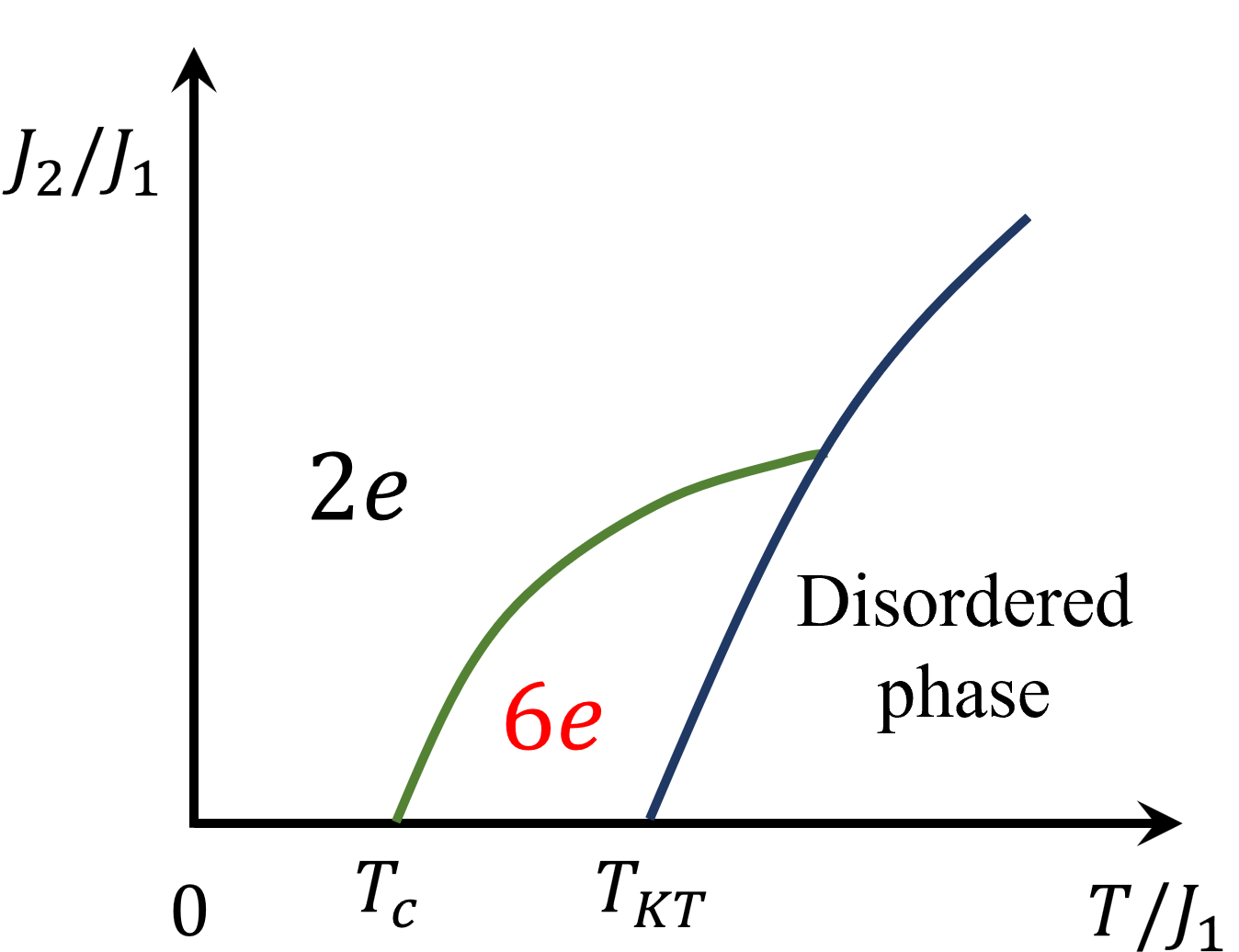}
\caption{A schematic phase diagram of the effective model in Eq.~(\ref{eq:chiralitymodel}) with respect to the $J_2/J_1$ and $T/J_1$.}
\label{fig:Figure_a5}
\end{figure}

The schematic phase diagram of the effective model Eq.~(\ref{eq:chiralitymodel}) is depicted in Fig.~\ref{fig:Figure_a5}.
In the regime of weak $J_2$, charge-$6e$ order emerges within an intermediate temperature regime, $T_{\mathrm{c}}<T<T_{KT}$.
As $J_2$ increases, $T_{\mathrm{c}}$ rises due to the enhanced energy cost resulting from it. 
The KT temperature $T_{KT}$ also rises with increasing $J_2$. 
This phenomenon aligns with the prediction of KT transition, where KT temperature is directly proportional to the superfluid stiffness \cite{kosterlitz1973ordering}.
In the model described by Eq.~(\ref{eq:chiralitymodel}), both the $J_1$-term and $J_2$-term contribute to superfluid stiffness in the low-energy theory, consequently leading to a larger $T_{KT}$ as $J_2$ increases when $J_1$ remains fixed.
It's important to acknowledge the limitations of Monte Carlo simulations in capturing the complexities of frustrated systems. 
A comprehensive establishment of the phase diagram is left for subsequent investigations, potentially employing advanced techniques such as state-of-the-art tensor-network methods.

%------------------------------------------------
%------------------------------------------------

\end{document}